\documentclass{PoS}
\usepackage{slashed}
\newcommand{\rd}{\mathrm{d}}
\newcommand{\PW}{\mathrm{W}}
\newcommand{\PZ}{\mathrm{Z}}
\newcommand{\Pg}{\mathrm{g}}
\newcommand{\alps}{\alpha_{\mathrm{s}}}
\newcommand{\rT}{\mathrm{T}}
\newcommand{\Or}{\mathcal{O}}

\title{Vector-boson production at the LHC:\\
 QCD and electroweak effects}

\ShortTitle{Vector-boson production at the LHC}

\author{\speaker{Tobias Kasprzik}\\
        Institut f\"ur Theoretische Teilchenphysik (TTP), \\
        Karlsruher Institut f\"ur Technologie (KIT)\\
        E-mail: \email{kasprzik@particle.uni-karlsruhe.de}}

      \abstract{A profound understanding of vector-boson production
        processes is of crucial importance at the LHC. The corresponding
        cross sections are large, and the final states are easy to
        reconstruct due to the clean signatures in the leptonic decay
        modes. Therefore, such processes play an important role as
        backgrounds in a large variety of new-physics signals, and they
        may furthermore help to better understand the well-established
        Standard-Model physics in a hadron-collider environment.  We
        review the recent progress in the theoretical description of
        higher-order QCD and electroweak effects in vector-boson
        production at the LHC, discussing the Drell--Yan
        process, vector-boson pair production and vector-boson
        production with associated jets.\\
\\
\\
{\it TTP11-27 \\
     SFB/CPP-11-57\\
     LPN11-59}
}

\FullConference{The 2011 Europhysics Conference on High Energy Physics, EPS-HEP
2011,\\
		July 21-27, 2011\\
		Grenoble, Rh\^one-Alpes, France}

\begin{document}

\section{The Drell--Yan process}\noindent
A proper understanding of the charged-current (CC) and neutral-current
(NC) single vector-boson production process $\mathrm{pp} \to
\PW^{\pm}/\PZ,\gamma^* \to l\nu_l / l\bar{l}$ is an important task at
the LHC. Such processes have a clean leptonic signature and are
therefore well suited for luminosity monitoring and
detector calibration purposes. Moreover, Z-boson production will allow
for a precise determination of the effective weak mixing angle at the
LHC by measuring appropriate forward-backward asymmetries, while W-boson
production is primarily used to precisely determine the W-boson mass
$M_{\PW}$ and width $\Gamma_{\PW}$, and to constrain the parton
distribution functions (PDFs) by analyzing the $\PW^-/\PW^+$-ratio or
the W-boson charge asymmetry $A_{\PW}^l \equiv (\rd \sigma_{l^+}/\rd
\eta_l -\rd \sigma_{l^-}/\rd \eta_l)/(\rd \sigma_{l^+}/\rd \eta_l +\rd
\sigma_{l^-}/\rd \eta_l)$, respectively.
\begin{figure}[]
\begin{center}
\includegraphics[width = 0.45\textwidth]{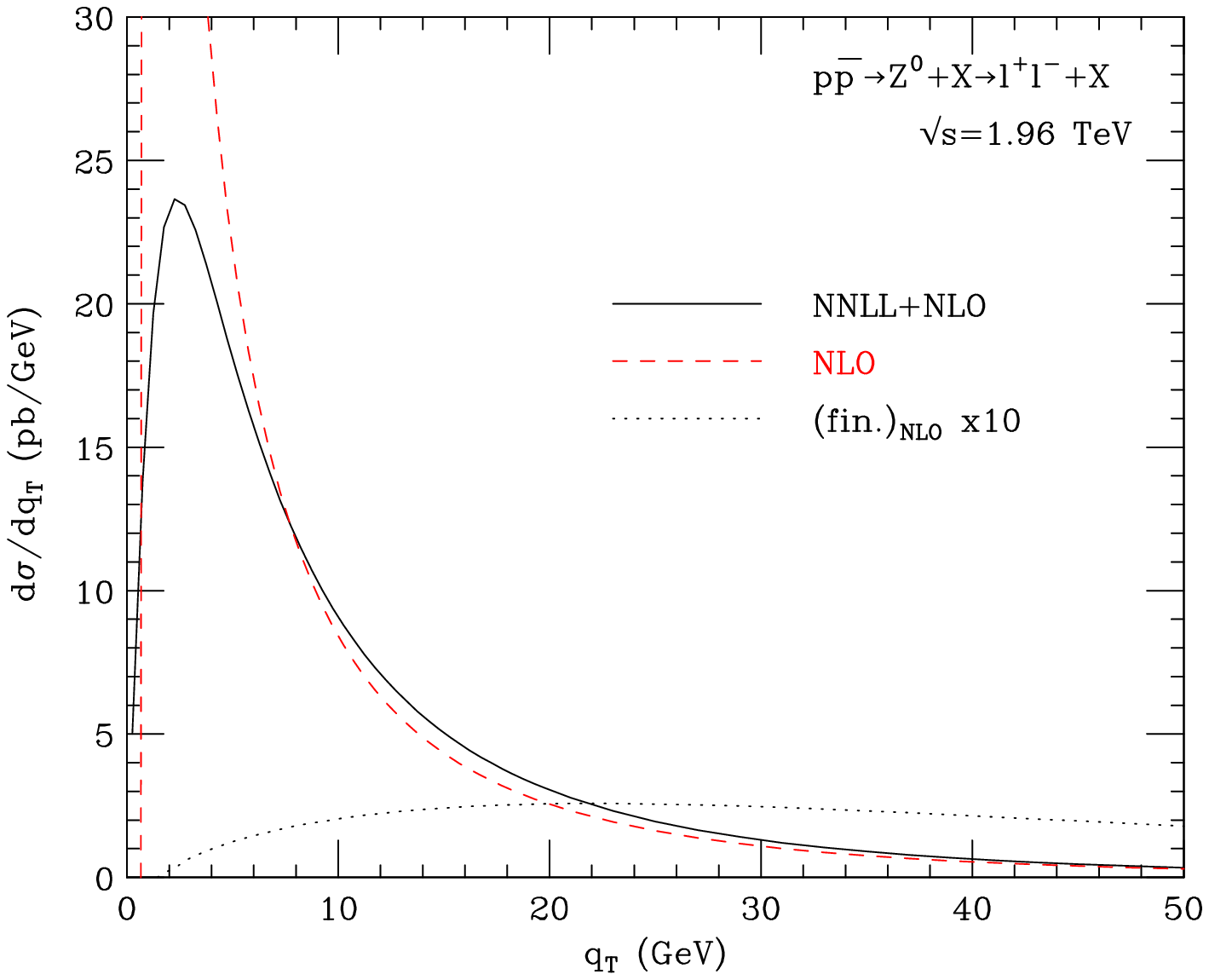}
\includegraphics[width = 0.5\textwidth]{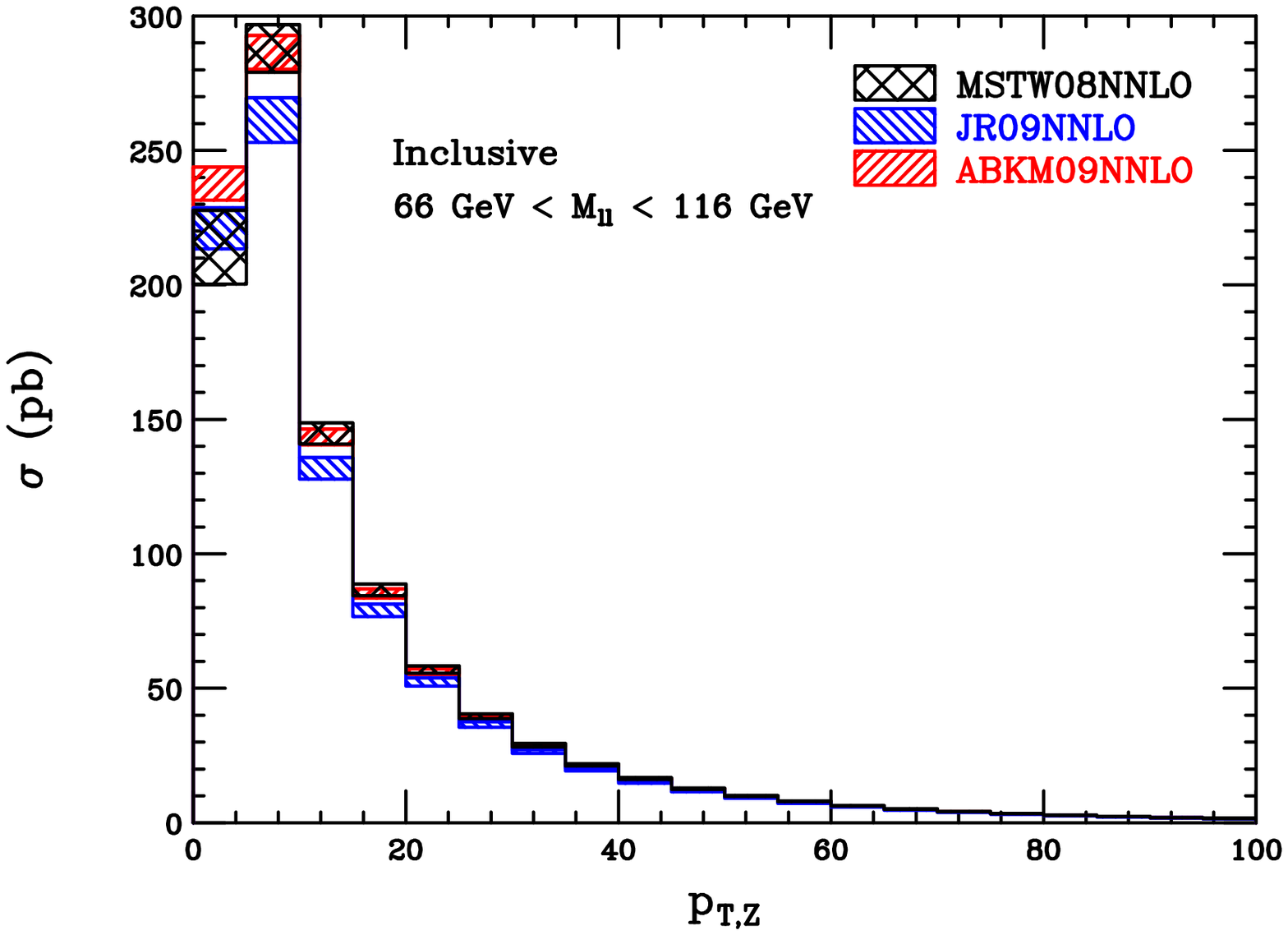}
\caption{Left: Comparison of the resummed result at NNLL accuracy to the
  fixed-order computation at the Tevatron~\cite{Bozzi:2010xn} for the
  Drell--Yan process. The fixed-order prediction fails for $q_{\rT} <
  10\, \mathrm{GeV}$. ($\mathrm{(fin.)_{NLO}}$:
  Non-logarithmically-enhanced contributions)\label{fi:DYRes}. Right:
  Distribution of the Z-boson transverse momentum at NNLO accuracy for
  different PDF sets at the LHC ({\tt FEWZ}~\cite{Gavin:2010az}).}
\end{center}
\end{figure}

At hadron colliders, at least the knowledge of the next-to-leading-order (NLO)
QCD corrections is crucial for the normalization of the total cross section,
the reduction of the scale dependence, and to obtain realistic predictions for
the shapes of differential cross sections. The QCD corrections to the
Drell--Yan (DY) process have been studied extensively. The NLO contributions
have been matched with parton
showers~\cite{Frixione:2006gn,Frixione:2007vw,Alioli:2008gx} and combined with
soft-gluon resummation~\cite{Arnold:1990yk, Bozzi:2010xn} to efficiently
account for dominating contributions $\propto \alps^n\ln^m(M_V^2/q_{\rT}^2)$
at small transverse momenta $q_{\rT}$ of the vector bosons (see
Fig.~\ref{fi:DYRes}, left). Moreover, the corresponding
next-to-next-to-leading-order (NNLO) two-loop corrections are known fully
differentially (see Fig.~\ref{fi:DYRes}, right) and have been implemented in
Monte Carlo programs~\cite{Melnikov:2006di, Catani:2009sm, Gavin:2010az}. Even
the $\mathrm{N^3LO}$ corrections are known in the soft-plus-virtual
approximation~\cite{Moch:2005ky}, pushing the theoretical uncertainties due to
perturbative QCD to the level of below 2\% for this specific process class.

The corresponding NLO electroweak (EW) corrections have been
investigated by many groups for the CC~\cite{Zykunov:2001mn,
  Dittmaier:2001ay, Baur:2004ig, CarloniCalame:2006zq, Brensing:2007qm}
and the NC~\cite{Baur:1997wa,CarloniCalame:2007cd,Dittmaier:2009cr}
process, and tuned comparisons of different implementations have been
performed.  Universal higher-order
corrections were included~\cite{CarloniCalame:2003ux}, and the
predictions were studied in different input schemes, where corrections 
due to $\Delta \alpha$ and $\Delta \rho$ are absorbed in effective
leading-order (LO) couplings. It was found that the relative EW
corrections are nearly insensitive to the specific theoretical treatment
of the vector-boson resonance and to the inclusion of virtual corrections
within the MSSM, respectively~\cite{Brensing:2007qm, Dittmaier:2009cr}.

\begin{figure}[]
\begin{center}
\includegraphics[width = .95\textwidth]{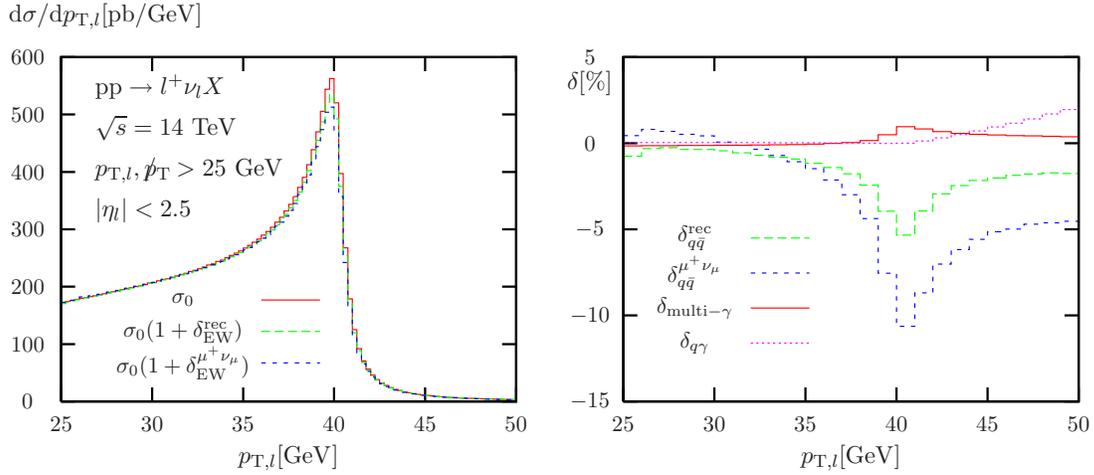}
\caption{EW corrections to the lepton transverse momentum for the CC
  Drell--Yan process at the LHC~\cite{Brensing:2007qm}. Results are
  presented for bare muons ($\delta_{q\bar{q}}^{\mu^+\nu_{\mu}}$) and
  for electrons employing electron-photon
  recombination ($\delta^{\mathrm{rec}}_{q\bar{q}}$), respectively; the
  effects due to multi-photon radiation
  ($\delta_{\mathrm{multi}-\gamma}$) as well as photon-induced processes
  ($\delta_{q\gamma}$) are small.\label{fi:EWCC}}
\end{center}
\end{figure}

In general, the EW corrections at moderate energies are dominated by
final-state photon radiation off leptons leading to a significant
distortion of the line-shapes of the leptonic invariant-mass and
transverse-momentum distributions (see Fig.~\ref{fi:EWCC}), which
strongly influences the precise determination of
$M_{\PW}$. Consequently, also the effect of multi-photon radiation has
been investigated in a structure-function
approach~\cite{Placzek:2003zg}, and the corresponding contributions have
been matched to the fixed-order $\mathcal{O}(\alpha)$ corrections within
{\tt HORACE}~\cite{Carloni Calame:2004qw}.  Moreover, first steps have
been taken towards a combined QCD $\otimes$ EW (two-loop)
analysis~\cite{Kilgore:2011pa} to improve the EW accuracy to the level
of better than 1\%.
\section{Vector-boson pair production}\noindent
\begin{figure}[]
\begin{center}
\includegraphics[width = 0.95\textwidth]{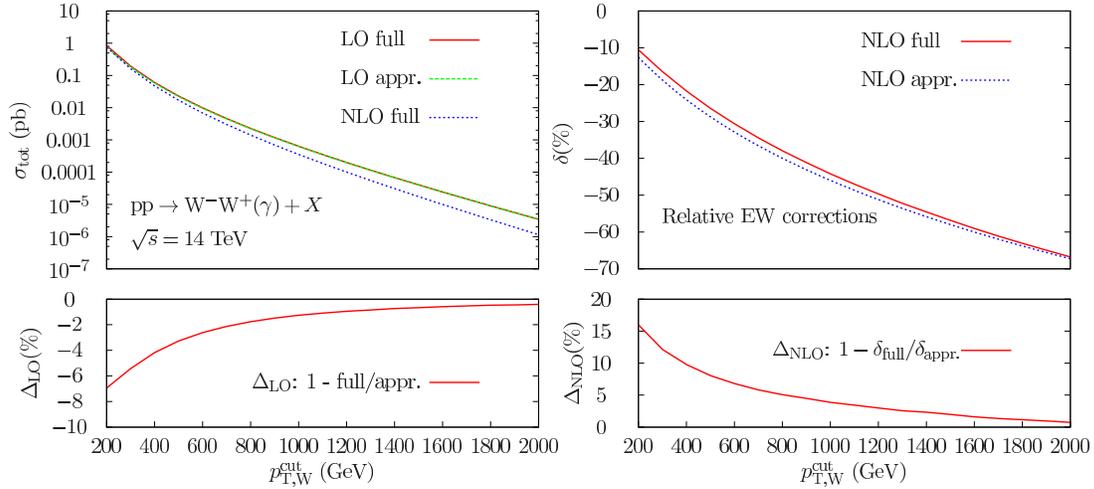}
\caption{W-boson pair production at the LHC at 14 TeV; Left: Comparison 
  of the full LO cross section to the high-energy approximation for
  different values of $p_{\rT,\PW}^{\mathrm{cut}}$. Right: Same for the
  relative corrections (Approximate NLO corrections at NNLL accuracy). For
  $p_{\rT,\PW}^{\mathrm{cut}} < 1\,\mathrm{TeV}$, the approximate results
  become crude~\cite{Bierweiler}.\label{fi:Wpair}}
\end{center}
\end{figure}
Vector-boson pair production processes $\mathrm{pp} \to V_1V_2 \to 4 l$
($V_i = \PW^{\pm},\PZ,\gamma$) play a key role in the understanding of
irreducible backgrounds to Standard-Model (SM) Higgs production in the
intermediate mass range. Moreover, they will allow us to probe the
non-abelian structure of the SM at high scales, and may give hints to
the existence of anomalous trilinear and quartic couplings, which are
predicted to have a sizable effect at high energies accessible at the
LHC. In addition, such processes constitute backgrounds to various
new-physics signatures with leptons and missing transverse energy.

The NLO QCD corrections to vector-boson pair production are known for a
long time, and a fully-exclusive computation of the two-loop corrections
to $\mathrm{pp} \to \gamma\gamma$ has been completed
recently~\cite{catani:ppgaga}. The NLO corrections have been matched
with parton showers and combined with soft-gluon
resummation~\cite{Frixione:2006gn,Nason:2006hfa}, and the leptonic
decays are accounted for in the narrow-width approximation, retaining
all spin information. Although the corresponding cross sections are
dominated by the $q\bar{q}$ channels, there are significant
contributions from the loop-induced channels $\Pg\Pg \to V_1V_2$ at the
LHC, especially if selection cuts for Higgs searches are
applied, of up to 30\%~\cite{Binoth:2006mf}.

In the high-energy limit the corrections at $\Or(\alpha)$ are known for
all combinations of external vector bosons in the pole
approximation~\cite{Accomando:2001fn} to facilitate a realistic
phenomenological description of the final-state leptons. Recently, we
have computed the full EW corrections to on-shell W-pair production at
the LHC~\cite{Bierweiler} and find sensible agreement with former
approximate results~\cite{Kuhn:2011mh} for center-off-mass energies
beyond 2~TeV (see Fig.~\ref{fi:Wpair}). For lower energies, however, the
approximation becomes crude even at leading order.  To this very
process, also two-loop pieces are available in NNLL
accuracy~\cite{Kuhn:2011mh}, leading to sizable positive corrections of
about 10\%.

\section{Vector-boson production with associated jets}\noindent
\begin{figure}
\begin{tabular}{cc}
\includegraphics[width=.45\textwidth]{Wm4jLHC7HTp_anti-kt-R5-Pt25_Allp_HT}
&
\includegraphics[width = .45\textwidth]{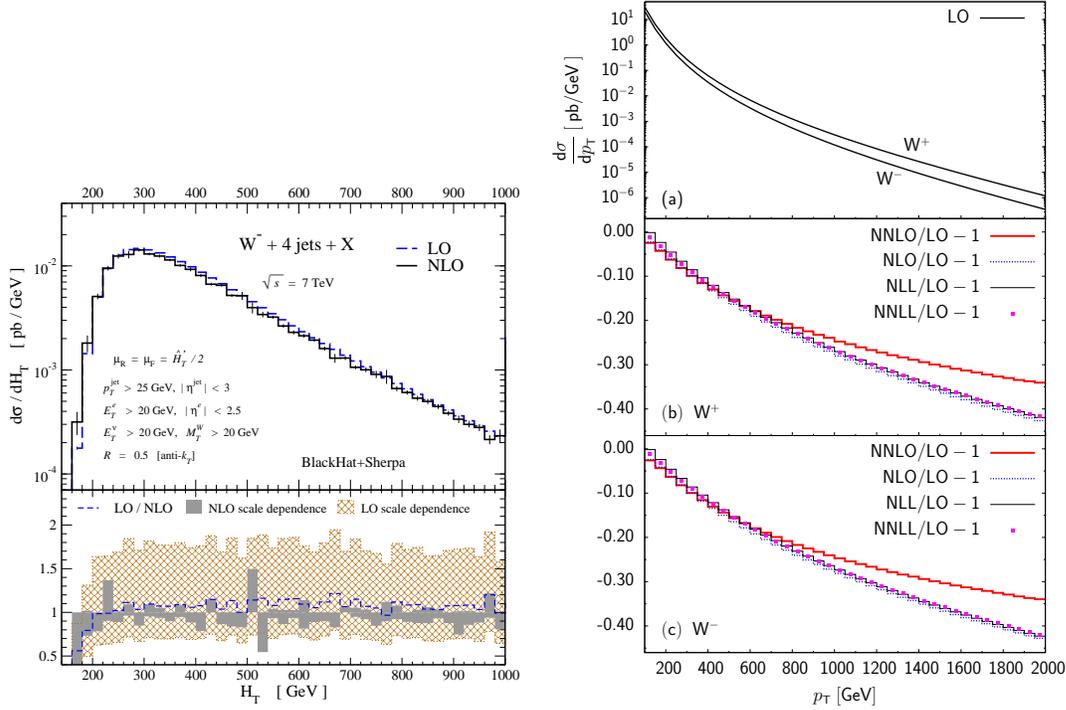}
\end{tabular}
\caption{\label{fi:Wjets} Left: NLO QCD corrections to $\PW^{-}$ + 4
  jets production at the LHC~\cite{Berger:2010zx}; top: Differential
  distribution of the transverse energy $H_{\rT} \equiv \sum_j
  E_{\rT,j}^{\mathrm{jet}} + E_{\rT}^e + E_{\rT}^\nu$; bottom:
  $K$-factor and scale dependence; note the marked reduction of scale
  uncertainties at NLO. Right: Relative EW corrections to the
  $p_{\rT}$ distribution for on-shell $\PW^{\pm}$ + jet production
  at the LHC, and corresponding relative
  corrections~\cite{Kuhn:2007cv}. (NNLO = NLO + 2-loop NLL; NLL and NNLL
  considered at one loop)}
\end{figure}
\noindent
At the LHC, vector bosons are almost always produced together with one
or more hard jets. Since the corresponding cross sections are still
sizable, and due to the distinct leptonic final states, such processes
may help to attain a profound understanding of SM physics in a
hadron-collider environment.  Moreover, V + jet(s) production also
provides a good possibility to study QCD jet dynamics in great detail,
and the corresponding signatures with jets, charged leptons and missing
transverse energy have to be understood properly to discriminate
hypothetical new-physics signals from the SM background.

The NLO QCD corrections for W/Z + 1 jet have been matched with parton
showers~\cite{Alioli:2010qp}, and the NLO corrections for W/Z + 2 jets
are e.g.\ included in {\tt MCFM}~\cite{Campbell:2002tg}. NLO results for
\mbox{W/Z + 3 jets} have been presented by different
collaborations~\cite{Berger:2010vm}, and even the NLO results for
$\PW+4\;\mathrm{jets}$ could be worked out~\cite{Berger:2010zx}, using
unitarity-based techniques for the automatized calculation of virtual
amplitudes (see Fig.~\ref{fi:Wjets}).

Various EW effects to W/Z + 1 jet production have been investigated,
both in the on-shell~\cite{Kuhn:2004em,Kuhn:2007cv} and
off-shell~\cite{Denner:2009gj} scenario, where the leptonic decays of
the vector bosons as well as finite-width effects are fully taken into
account in the latter. As in the DY case, at high energies the
corrections are dominated by universal large logarithms $\propto
\alpha^L\ln^{2L-n}(M_V/\sqrt{s})$ ($\equiv$ N$^n$LL accuracy at $L$
loops) which are known up to two loops in the NLL approximation (see
Fig.~\ref{fi:Wjets}), while at low and medium energies the corrections
are nearly fully dominated by final-state photon radiation.

\section{Conclusions}\noindent
Precise theoretical predictions for vector-boson production at the LHC
have been established during the recent years, where both fixed-order
and resummed QCD and EW effects were considered. The accurate knowledge
of Drell--Yan physics will help to further constrain the PDFs and
probably enable the most accurate determination of $M_{\PW}$ with an
error of only 15 MeV at the LHC. Moreover, it will also open the door to a
fruitful analysis of EW precision observables at hadron
colliders. Apart from that, a profound theoretical knowledge of
vector-boson pair production and V + jet(s) is mandatory to properly
assess the SM backgrounds to Higgs-boson or $\mathrm{t\bar{t}}$
production, as well as to various beyond-SM physics scenarios.

\end{document}